\documentclass[aps,prl,twocolumn,superscriptaddress,showpacs,floatfix]{revtex4}
\usepackage{graphicx,epsfig}
\usepackage{amssymb,times}
\usepackage{amsmath,amsfonts,bm}
\usepackage{grffile}
\usepackage{dcolumn}
\graphicspath{{plots/}}

\def\beq{\begin{equation}}
\def\eeq{\end{equation}}

\newcommand{\ubc}{Department of Physics and Astronomy, University of British Columbia, Vancouver V6T 1Z1, Canada.}
\newcommand{\bonn}{HISKP, University of Bonn, Nussallee 14-16, D-53115 Bonn, Germany.}
\newcommand{\salerno}{CNR-SPIN and Dipartimento di Fisica ``E. R. Caianiello'', Universit\`a degli Studi di Salerno,
Via Giovanni Paolo II, I-84084 Fisciano, Italy.}
\newcommand{\lyon}{Laboratoire de Physique de l'\'Ecole Normale Sup\'erieure de Lyon,
CNRS UMR5672, 46 All\'ee d'Italie, F-69364 Lyon Cedex 7, France.}

\begin{document}

\title{Correlation dynamics during a slow interaction quench in a one-dimensional Bose gas}
\author{Jean-S\'ebastien Bernier}
\affiliation{\ubc}
\author{Roberta Citro}
\affiliation{\salerno}
\author{Corinna Kollath}
\affiliation{\bonn}
\author{Edmond Orignac}
\affiliation{\lyon}

\begin{abstract}
We investigate the response of a one-dimensional Bose gas
to a slow increase of its interaction strength. We focus on the rich dynamics of equal-time single-particle correlations
treating the Lieb-Liniger model within a bosonization approach and the Bose-Hubbard model using the time-dependent
density-matrix renormalization group method. For short distances, correlations follow a power-law with distance with an
exponent given by the adiabatic approximation. In contrast, for long distances, correlations decay algebraically with an exponent understood
within the sudden quench approximation. 
This long distance regime is separated from an intermediate distance one by a generalized Lieb-Robinson criterion. 
At long times, in this intermediate regime, bosonization predicts that single-particle correlations decay following a stretched exponential. 
This latter regime is unconventional as, for one-dimensional interacting systems, the decay of single-particle correlations is usually
algebraic within the Luttinger liquid picture. 
We develop here an intuitive understanding for the propagation of correlations, in terms of a generalized light-cone, 
applicable to a large variety of systems and quench forms.

\end{abstract}

\date{\today}

 
\maketitle

{\it Introduction:} Recent advances in the development of fast probing and control techniques applicable to correlated systems
have opened up the possibility to dynamically prepare complex quantum many-body states.
For example, effective phase transitions have been induced through the application of external driving
fields~\cite{LignierArimondo2007,BasovHaule2011,StruckSengstock2011} and
states, such as a Bell state of ions or a Tonks-like state in a quantum gas, have been realized
using tailored environments~\cite{BarreiroBlatt2010,SyassenDuerr2008}. In fact, the
dynamical preparation of states promises to have an important impact in fields as diverse as condensed
matter physics, quantum information, quantum optics and ultracold atomic physics.
On the theoretical side, despite tremendous progress in recent years, many of the basic concepts behind
the dynamical generation of states still remain to be understood. 

In this article, we focus on the preparation of unconventional states in isolated systems 
using slow parameter changes. Considerable
experimental efforts have been devoted to understand slow quench 
dynamics~\cite{greiner_2002,HungChin2010,sherson_2010,bakr_2010,demarco_2011}. 
However, in these works, as well as in many theoretical ones (see Ref.~\onlinecite{polkovnikov_review} and references therein),
the emphasis has been put on understanding how energy is absorbed and defects produced.

In recent years, the focus has partially shifted towards the study of longer
range correlation dynamics during a slow parameter
quench~\cite{SchuetzholdFischer2006,CucchiettiZurek2007,CincioZurek2007,CherngLevitov2005,PolkovnikovGritsev2008,
EcksteinKollar2009,MoeckelKehrein2010,DoraZarand2011,DziarmagaTylutki2011,PolettiKollath2011,BernierKollath2012}.
Understanding the evolution of such correlations is paramount as the nature of many-body quantum states are typically
characterized by longer range correlators. Interestingly, light-cone-like spreading~\cite{LiebRobinson1972,CalabreseCardy2006} 
of parity correlations, both in space and time, has even been
observed experimentally in an interacting one-dimensional bosonic gas after a sudden quench of the optical lattice depth~\cite{CheneauKuhr2012}.
For slow quenches, a similar linear light-cone-like evolution of correlations has been predicted for density correlations
in bosonic systems~\cite{DziarmagaTylutki2011} and for single-particle correlations in fermionic systems~\cite{DoraZarand2011}.

We analyze here the correlation dynamics during a slow linear increase of the interaction strength,
at zero temperature, in two paradigmatic one-dimensional interacting models: 
the Lieb-Liniger and Bose-Hubbard models. 
We show that
a generalized Lieb-Robinson bound describes the evolution of single-particle correlations.
This bound can be understood within a simple picture involving quasiparticle pairs created during the quench. At each
instant in time, the quasiparticles propagate at their instantaneous velocity: as this velocity is time-dependent the
evolution front possesses a non-trivial functional form. 
This non-trivial form contrasts with the linear evolution front of correlations, the horizon, which 
arises after a sudden quench and is due to a constant quasiparticle velocity~\cite{CalabreseCardy2006}.
The structure of the correlation front can be
extracted solely from the knowledge of the quasiparticle velocity and does not require a detailed understanding 
of the more complicated correlation function. In fact, the approach developed here can be applied
to various interacting systems.

For the one-dimensional models under study in this article, we find that outside the bound
the single-particle correlations decay algebraically 
with distance 
with an exponent determined by the initial Luttinger parameter 
and decreased amplitude.
In contrast, inside the bound, the correlations present much more interesting dynamics. For short distances, the
algebraic decay depends on the ramp time~\cite{PolkovnikovGritsev2008,DoraZarand2011}. While for larger distances 
and quench times, the correlations, within the Lieb-Liniger model, decay following a stretched exponential. 
This particular decay form is unexpected as, even for instantaneous quenches, an algebraic decay persists at all distances and
times~\cite{cazalilla_quench}. 
A similar stretched exponential behavior was found in Ref.~\onlinecite{PolkovnikovGritsev2008} (without a
time-dependent prefactor).
In the rest of the article, we analyze in detail the evolution of
single-particle correlations, and highlight the different regimes both in position and momentum space.


{\it Model:} Bosonic atoms in a one-dimensional wave guide can be described by the Lieb-Liniger (LL) model
\begin{eqnarray}
\label{eq:lieb-liniger}
H &=& \int dx \left[ -\frac{\hbar^2}{2m} \psi^\dagger(x) \partial_x^2 \psi(x) + \frac {g(t)} 2 \rho(x)^2 \right]
\end{eqnarray}
with $\psi(x)$ the boson annihilation operator and $\rho=\psi(x)^\dagger
\psi(x)$ the density. The interaction strength $g$ is related to the
$s$-wave scattering length $a_s$ of the atoms and to the transverse
trapping frequency $\omega_\perp$ by $g\approx 2\pi \hbar \omega_\perp a_s$.
We assume that the gas is initially prepared at a certain interaction strength
$g(t)=g_0$ and that for $t>0$ a linear variation of the interaction strength of
the form $g(t)=g_0+(g_f-g_0)\frac{t}{t_f}$ is performed. Experimentally this
variation can be achieved, for example, by using a Feshbach
resonance or by varying the intensity of the transverse trapping~\cite{BlochZwerger2008}.

A similar interaction quench can be done by confining bosonic atoms to an optical lattice potential along
the one-dimensional direction. The theoretical model describing this situation is the Bose-Hubbard model given by
\begin{eqnarray}
\mathcal{H} &=& -J \sum_l \left(b^{\dagger}_{l+1} b_l + \text{h.c.}\right) + \frac{U(t)}{2} \sum_l \hat{n}_l(\hat{n}_l-1) \nonumber
\end{eqnarray}
with $b^{\dagger}_l$ the operator creating a boson at site $l$ and $\hat{n}_l=b^{\dag}_l b_l$ the local density operator.
The first term of the Hamiltonian corresponds to the kinetic energy of atoms with hopping amplitude $J$
while the second term is the potential energy with onsite interaction of strength $U$.
Taking the continuum limit of the Bose-Hubbard model in the superfluid phase, this model can be mapped onto the
LL Hamiltonian~\cite{KollathZwerger2003}. In this case, the linear interaction quench $g(t)$ translates into a linear change
of the interaction amplitude $U(t)$.

For both models, in the superfluid phase, the low energy physics is well described by the
Tomonaga-Luttinger liquid (TLL) Hamiltonian~\cite{giamarchi_book_1d, Cazalilla2011}
\begin{eqnarray}
\label{eq:tll-noneq-fourier}
H \!\!&=&\!\! \sum_q \frac{q^2}{2\pi} \left[ u(t)K(t)  \theta(q)\theta(-q) +\frac{u(t)}{K(t)} \phi(q)\phi(-q)\right]\,\,\,\,\,\,
\end{eqnarray}
where
$\phi(x)=\frac{1}{\sqrt{L}}\sum_q \phi(q) e^{iq x} e^{-|q|\alpha/2}$ and
$\theta(x)=\frac{1}{\sqrt{L}}\sum_q \theta(q) e^{iq x}e^{-|q|\alpha/2}$ are conjugate fields
satisfying the canonical commutation relation $\lbrack \phi(x), \nabla \theta(x')\rbrack=i \pi \delta
(x-x')$. We have set here $\hbar = 1$ and $\alpha$ is a short distance cut-off.
The sound velocity $u$ and the Luttinger parameter $K$ are related to the parameters of
the original Hamiltonians. These parameters can, for example, be extracted from the
Bethe Ansatz solution of Eq.~(\ref{eq:lieb-liniger})~\cite{lieb_bosons_1D} or through
numerical approaches for the Bose-Hubbard Hamiltonian~\cite{kuhner_bose_hubbard_critical_point,KollathZwerger2003}.

In the LL model, the Galilean invariance ensures that the product $u(t)K(t)$
remains unchanged upon varying the interaction parameters~\cite{haldane_bosons} and thus $u(t)K(t)=u_0K_0$. For small linear
changes of these parameters, this translates, to first order in the variation, to a time-dependent
ratio $\frac{u(t)}{K(t)} \approx \frac{u_0}{K_0} (1+\frac{t}{t_0})$ with
$t_0=\frac{\pi u_0 t_f}{K_0 (g_f-g_0)}$ and a typical lengthscale $l_0=u_0t_0$. This result is then used to obtain expressions for the time-dependent
sound velocity $u(t)\approx u_0\sqrt{1+\frac{t}{t_0}}$ and for the time-dependent Luttinger
parameter $K(t) \approx K_0/\sqrt{1+\frac{t}{t_0}}$. These expressions are still valid for small parameter variations
in the Bose-Hubbard model given the relation $U_\text{0/f}~a = g_\text{0/f}$ where $a$ is the lattice constant.

A major distinctive feature of the TTL model is that its low energy excitations are collective modes
(density fluctuations) instead of individual quasiparticles. Hence, only quasi-long range order
persists even down to zero temperature. This situation is exemplified by the anomalous (non-integer) power-law
dependence of its correlation functions~\cite{giamarchi_book_1d}. Moreover, the time-dependence does
not introduce couplings between the different momentum modes of the TTL Hamiltonian.
This leads to momentum decoupled equations of motion for the Fourier components of the fields of 
the form~\cite{DoraZarand2011,DziarmagaTylutki2011, PolkovnikovGritsev2008}
\begin{eqnarray}
\label{eq:eoms}
\frac{d}{dt} \phi(q) = u_0K_0~q~\theta(q)~~~\text{and}~~~
\frac{d}{dt} \theta(q) = -\frac{u(t)}{K(t)}~q~\phi(q).
\end{eqnarray}
The solutions for these equations of motion can be written using bosonic quasiparticles with
creation and annihilation operators $a^\dagger$ and $a$ which diagonalize the Hamiltonian at $t=0$:
\begin{eqnarray}
\label{eq:eom-solutions}
\phi(q,t)\! &=& \! 2~\sqrt{\pi K_0}~|q| \left[ a_q F^* + a^\dagger_{-q} F \right],\\
\theta(q,t)\! &=&\! \frac 1 {u(t) K(t)q} \sqrt{\frac{\pi K_0}{2|q|}}\left[a_q \frac{d}{dt}F^* + a^\dagger_{-q} \frac{d}{dt}F \right]
\end{eqnarray}
where $F(q,t)$ is the solution of the equation
\begin{eqnarray}
\left(\frac 1 {u_0 K_0} \frac{d^2}{dt^2} F(q,t) \right) = -\frac{u(t)}{K(t)}~q^2~F(q,t) \nonumber
\end{eqnarray}
with initial conditions $F(q,0)=1$, $\frac{d}{dt}F(q,t)|_{t=0}=i u_0 |q|$.
This solution can be expressed in terms of Bessel functions
(see Eq.~(9.1.51) of Ref.~\onlinecite{abramowitz_math_functions}):
\begin{eqnarray}
\label{eq:bessel-ramp}
F(q,t)\!\!\!&=&\!\!\!\frac{\pi s~\tau^{\frac{1}{2}}}{\sqrt{3}}\left[J_\frac{2}{3}\left(s\right)  J_\frac{1}{3}(s \tau^\frac{3}{2})
+  J_{-\frac{2}{3}}\left(s \right)  J_{-\frac{1}{3}}(s \tau^\frac{3}{2}) \right. \nonumber \\
&&\!\!+i \left. \left(J_{-\frac{1}{3}}\left(s\right)  J_\frac{1}{3}(s \tau^\frac{3}{2})- J_\frac{1}{3}\left(s\right)
J_{-\frac{1}{3}}(s\tau^\frac{3}{2})\right)\right]
\end{eqnarray}
where $s(q,t_0)=\frac 2 3 l_0 |q| $ and $\tau(t,t_0)=1+\frac{t}{t_0}$ are the dimensionless momentum and 
time, respectively~\cite{footnote}.

{\it Evolution of the single-particle correlation function:}
In the following, we survey the rich behavior of the equal-time single-particle correlation
function $G(x,t) =\frac 1 2 \langle \psi(x,t) \psi^\dagger(0,t) +h.c.\rangle$
during a slow interaction quench.

In the bosonization representation, the equal-time single-particle correlation function takes the form
\begin{eqnarray}
\label{eq:green-2}
G(x,t)_{q\simeq 0} &=& A_0^2  \langle e^{i\theta(x,t)}  e^{-i\theta(0,t)}\rangle \nonumber \\
&=& A_0^2 e^{-\frac{1}{2} I(\xi,\tau,\tilde{\alpha})}
\end{eqnarray}
where $A_0$ is a non-universal constant which depends on the underlying
microscopic model. We introduced for convenience the dimensionless length $\xi=\frac{3x}{2l_0}$ and,
correspondingly, the dimensionless short distance cut-off $\tilde{\alpha}=\frac{3\alpha}{2l_0}$.
The function $I(\xi, \tau, \tilde{\alpha})$ of Eq.~(\ref{eq:green-2}) is then given by
\begin{eqnarray}
\label{eq:theta-correlator}
I(\xi, \tau, \tilde{\alpha}) &=& \frac{\pi^2 \tau^2}{3 K_0} \int_0^\infty ds~s~e^{-\tilde{\alpha}s}~(1-\cos s\xi) \\
&&\times~ \left[ \left(J_\frac{2}{3}(s) J_{-\frac{2}{3}}(s\tau^\frac{3}{2}) - J_{-\frac{2}{3}}(s) J_\frac{2}{3}(s \tau^\frac{3}{2})\right)^2 + \right. \nonumber \\
&&~~~~ \left. \left(J_{-\frac{1}{3}}(s) J_{-\frac{2}{3}}(s \tau^\frac{3}{2}) + J_\frac{1}{3}(s) J_\frac{2}{3}(s \tau^\frac{3}{2})\right)^2 \right]. \nonumber
\end{eqnarray}
From the equation above it immediately follows that
Eq.~(\ref{eq:green-2}) only depends on the dimensionless variables
$\tau$, $\xi$, $\tilde{\alpha}$ and not separately on $t$, $t_0$, $x$ and $\alpha$.
This implies that, for a given final value of the interaction strength, increasing the ramp velocity, $\frac{1}{t_0}$,
mainly enters the expressions through an increased rescaled length $\xi$.

\begin{figure}[h!]
\centering
\includegraphics[width=\linewidth]{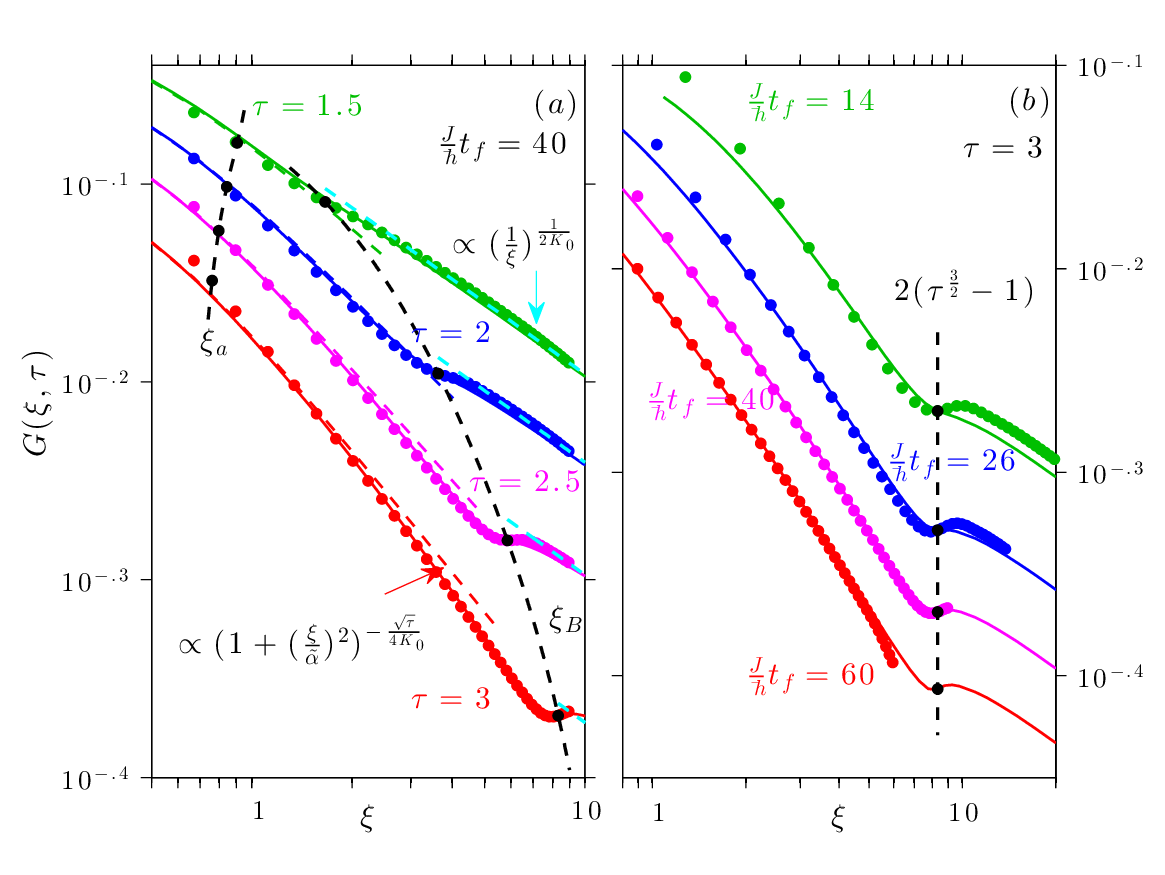}
\caption{Decay of single-particle correlations with increasing distance for different $\tau$ and $t_f$. Comparison
between results obtained using bosonization Eq.~(\ref{eq:green-2}) with Luttinger liquid parameters
$K_0 = 4.1561$ and $u_0 = 1.3323$ (solid lines) and using time-dependent density-matrix
renormalization group (t-DMRG) for the Bose-Hubbard model (circles) for a quench from $U_0 = J$
(lattice length: $L = 100$, filling: $n = 1$, maximum number of bosons per site: $6$).
$(a)$ Time evolution for different values of $\tau$ and for a fixed value of $t_f = 40 \frac{\hbar}{J}$.
The two dashed lines intersecting all $\tau$ data sets are the bounds: (left) $\xi_a = \tau^{-1/4}$
and (right) $\xi_B = 2 (\tau^{3/2} - 1)$. The colored dashed lines on the left of $\xi_B$ 
are curves proportional to the function $(1 + (\frac{\xi}{\tilde{\alpha}})^2)^{-1/(4K(\tau))}$; while
the dashed lines on the right of $\xi_B$ are curves proportional to the function $\xi^{-1/(2K_0)}$.
$(b)$ Comparison between different ramp times $t_f$ for a fixed value of $\tau=3$. The vertical dashed
line is the bound $\xi_B = 2 (\tau^{3/2} - 1)$.}
\label{Fig:comparison}
\end{figure}

{\it Asymptotic expansion of the single-particle correlation function:}
The time evolution of single-particle correlations described by Eqs.~(\ref{eq:green-2}) and (\ref{eq:theta-correlator})
is extremely rich. Typical time evolutions of these correlations with distance are shown in panel $(a)$ of
Fig.~\ref{Fig:comparison} for both the Bose-Hubbard model and the bosonization approach.
For the chosen parameters, we found very good agreement between the two evolutions at longer distances, as long as an additional
time-dependent prefactor is multiplied to the expression obtained using bosonization.
This prefactor corrects for the short distance behavior which is not properly taken into account by the low energy
theory. As expected, the bosonization description works best for slow and small parameter changes. In particular, deviations
are observed when the Mott-insulating phase of the Bose-Hubbard model is approached or when too many excitations are
created.

Initially, before the slow quench begins (at $\tau = 1$ within our formalism), the correlation function decays
algebraically with distance as
$G(\xi) = A_0^2~(1+ (\xi/\tilde{\alpha})^2)^{-1/(4K_0)}$.
This behavior is typical of a Luttinger liquid. Then, as the interaction strength is slowly ramped up, the
form of the correlation function evolves. For small $\xi$ and sufficiently short $\tau$,
changes are minimal as the correlation function still
decays algebraically, but the exponent is now determined by the time-dependent
Luttinger parameter $K(t) = K_0/\sqrt{\tau}$ showing up in the exponent~\cite{supplofpaper}.
This result implies that for short dimensionless distances, $\xi_a := \tau^{-1/4} \gg \xi$, the correlations react instantaneously
to the slow interaction change and adjust to the ground state decay corresponding to the current
interaction value (see panel $(a)$ of Fig.~\ref{Fig:comparison}). The main contribution to this mechanism comes
from quasiparticles with large momenta $q \gg \frac{1}{l_0}$. This adiabatic regime spatially decreases with time and disappears
completely when $\xi_a(t)\approx \tilde{\alpha}$, where $\tilde{\alpha}$ is the dimensionless short distance cut-off.

For larger distances, the correlations deviate much more from their standard initial form and a 
dip appears. The formation of this dip is a clear signal of the non-equilibrium nature of the physics at play.
For distances beyond this dip, the initial algebraic decay, $\xi^{-1/(2K_0)}$, reappears as one can see in panel $(a)$ of Fig.~\ref{Fig:comparison}.
The position of the dip coincides approximately with the correlation evolution front.
The time-dependent position of this front can be understood 
by considering the propagation of quasiparticles.
At any given time $t$, the system Hamiltonian is diagonal in its
instantaneous quasiparticles as $H(t') =  \sum_q u(t') |q| a_q^\dagger(t') a_q(t')+\frac{1}{2}$.
Assuming discrete time steps, this means that the action of the Hamiltonian at time $t-\delta t$, diagonal
in its own quasiparticles, has created (and annihilated) entangled quasiparticle pairs $a^\dagger_q(t) a^\dagger_{-q}(t)$.
These entangled quasiparticles, forming a pair, propagate with velocity $u(t)$
in opposite direction and thereby carry correlations over a distance $2~u(t)~dt$ within a time interval $dt$.
Hence, for points separated by a distance $\xi$ larger than $\xi_{B} =  \frac{3}{l_0} \int_0^t dt'~u(t')$, the single-particle 
correlation decay is unaffected by the change in the interaction aside from an overall prefactor. 
For the system under study, $u(t) = u_0 \sqrt{1+\frac{t}{t_0}}$ and we find that
$\xi_{B} = 2~(\tau^{3/2}-1)$.  Thus, the evolution front beyond which correlations still follow the initial algebraic decay is
given by $\xi_B$ as evidenced in Fig.~\ref{Fig:comparison}. In particular, the position of the bound
does not depend on the ramp velocity and time separately as can be seen in panel $(b)$ of Fig~\ref{Fig:comparison}.
One clearly sees from there that, for a given $\tau$, the position $\xi$ of the dip (measured in units of $l_0$) is
the same for different ramp times. The existence of such a propagation front is reminiscent of the
light-cone-like evolution of correlations recently investigated in the context of instantaneous
quenches~\cite{LiebRobinson1972,bravyi2006,CalabreseCardy2006,LaeuchliKollath2008,CheneauKuhr2012,Mitra2013}.

\begin{figure}[h!]
\centering
\includegraphics[width=\linewidth]{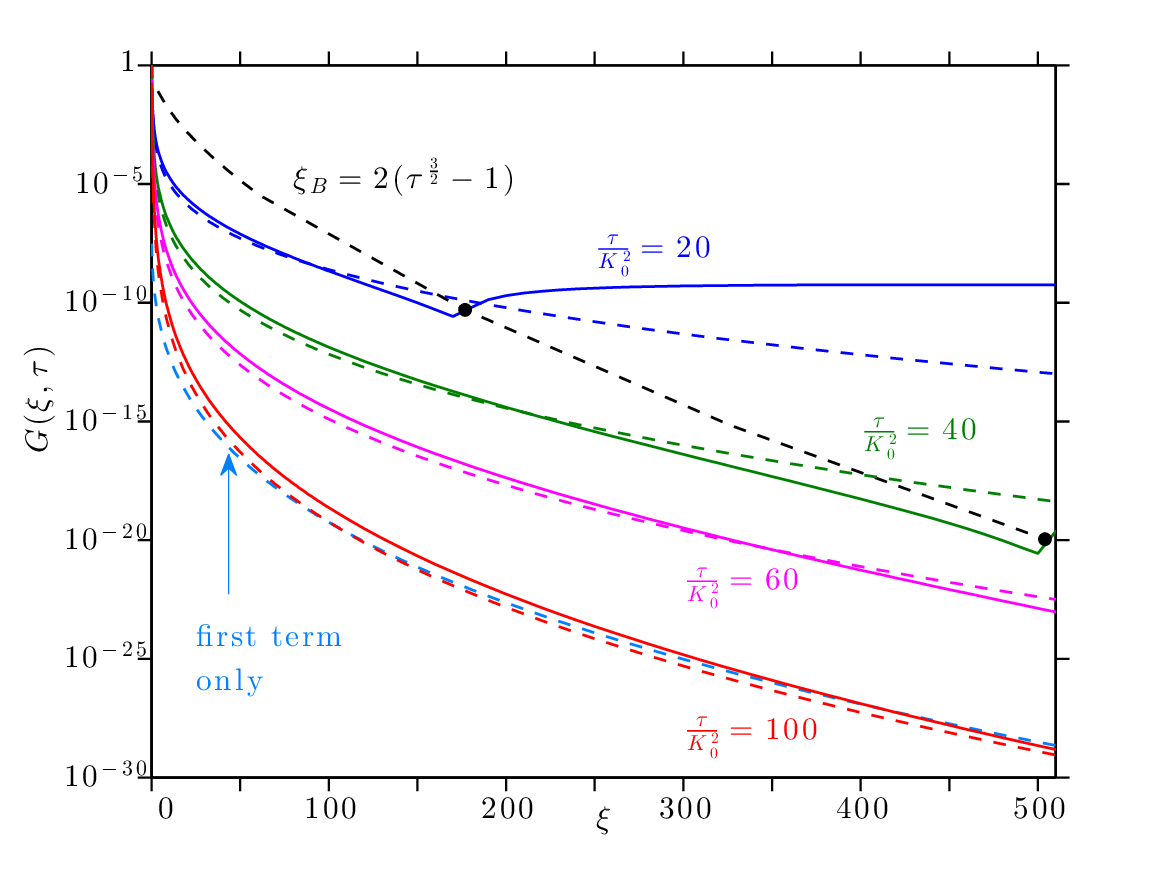}
\caption{Behavior of single-particle correlations with increasing distance for large values of $\tau/K_0^2$.
Exact evaluation of the bosonization expression, Eq.~(\ref{eq:theta-correlator}) -- solid lines, is compared to the full
approximate expression, Eq.~(\ref{eq:stretched_exp_full}) -- dashed lines. For $\tau/K_0^2 = 100$, we also compare the
exact expression to the first exponential term of Eq.~(\ref{eq:stretched_exp_full}). In the large $\tau$ limit, if one adjusts
the prefactor correctly, the stretched exponential provides a good description of the correlation decay before $\xi_B$.
The black dashed line indicates the position of the evolution front $\xi_B = 2~(\tau^{3/2}-1)$.
Used parameters: $s_\text{min} = 10^{-5}$, $s_\text{max} = 60$ (the lower and upper cut-offs 
in Eq.~(\ref{eq:theta-correlator})) and $\tilde{\alpha} = 0.1$.}
\label{Fig:stretched}
\end{figure}

For larger dimensionless times, as illustrated in Fig.~\ref{Fig:stretched}, an additional decay regime takes place at
intermediate distances before the bound $\xi_B$. This interesting behavior shows up in the bosonization approach and takes the form
\begin{eqnarray}
\label{eq:stretched_exp_full}
&& G(\xi, \tau)_{q \simeq 0} \simeq \tilde{\mathcal{C}}(\tau) \times \\
&&~~~~~~ \exp\left(-\frac{2^{\frac{1}{3}} \pi^2 \tau^{\frac{1}{2}}}{K_0 \Gamma(\frac{1}{3})^3}~\xi^{\frac{1}{3}}\right)
\exp\left(\frac{\pi^{\frac{3}{2}} \tau^{\frac{1}{2}} \Gamma(\frac 1 6)}
{6 K_0 \Gamma(\frac 1 3) \Gamma(\frac{2}{3})^2} \frac{1}{\xi^{\frac{1}{3}}} \right) \nonumber
\end{eqnarray}
with $\tilde{\mathcal{C}}(\tau)$ a prefactor independent of $\xi$. For intermediate $\tau$, both exponential terms are
required to adequately reproduce the behavior of Eq.~(\ref{eq:theta-correlator}) as shown in
Fig.~\ref{Fig:stretched}. However, for values of $\tau$ whose corresponding bound $\xi_B$ is located at sufficient large $\xi$,
only the first exponential term is important. In this case single-particle correlations
decay with distance as a stretched exponential, a similar decay was found in Ref.~\onlinecite{PolkovnikovGritsev2008}. 
Such a functional form is unconventional for Luttinger liquids as, typically, correlations decay algebraically in these systems. Even for
sudden interaction quenches in both bosonic and fermionic systems~\cite{cazalilla_quench, KarraschMeden2012} and for slow 
quenches in fermionic systems~\cite{DoraZarand2011},
only algebraic decay of correlations have been uncovered.
The presence of such an unusual functional form is mainly due to
the reinforcement of the amplitude of phase
fluctuations at low momenta with respect to the equilibrium case.
The quench generates an unusual (non-thermal) distribution of quasiparticles around 
$\frac{3}{2\xi_B~l_0} < q \le \frac{1}{l_0}$~\cite{supplofpaper}.

Moreover, as the appearance of the stretched exponential decay is limited to large values of $\tau$, this regime only occurs for
relatively large parameter changes $t \gg t_0$.
It is still an open question, whether this stretched exponential decay regime arises within
the Bose-Hubbard model. As this regime only occurs for large parameter changes, the TLL model might
not describe properly the dynamics of the Bose-Hubbard model and relaxation mechanisms not present in the
TLL model might dominate the evolution. A careful analysis of this
last point would be extremely valuable but is left to further studies.

{\it Experimental implementation and detection:} 
One-dimensional interacting bosonic gases have been realized experimentally
using various setups~\cite{StoeferleEsslinger2004,ParedesBloch2004,KinoshitaWeiss2004}.
The time-dependence of the ratio of potential to kinetic energy can be implemented using Fesbach resonances, or by
varying the optical lattice depth or the transverse trapping.

Detection of the single-particle correlation function can also be carried out
experimentally. Using radio-frequency pulses, atoms can be outcoupled from the one-dimensional Bose gas
at two spatially separated positions and their interference is
then observed after a free fall.
This technique was successfully employed to measure the build-up of equal-time single-particle
correlations in a Bose-Einstein condensate after a sudden decrease of its temperature~\cite{RitterEsslinger2006, DonnerEsslinger2007}.
Another possible detection scheme
relies on time-of-flight (TOF) measurements which provide, in the far-field limit,
access to the momentum distribution $n(q)=\int dx~e^{iqx}~G(x)$. The very long distance behavior of the single-particle
correlation is
dominated by the Luttinger liquid power-law;
however, at a critical wavevector, $q_c$,
determined by the ballistic expansion condition, a crossover occurs and $n(q)$ is dominated by the
Fourier transform of the stretched exponential.
Therefore, at $q_c \sim m_B \xi_B l_0/t$ (with $m_B$ the atom mass) a crossover should be visible in the TOF measurements. 
One of the main challenges towards the observation of the evolution of correlations will be the realization of
a relatively homogeneous gas as inhomogeneities can cause mass transport and mask the
internal evolution~\cite{NatuMueller2011,BernierKollath2012}. 
However, due to recent experimental advances~\cite{bakr_2010, GauntHadzibabic2013}, we believe
that, in the future, creating approximately box-shaped one-dimensional gases will be possible.

{\it Conclusion:}
We uncovered various interesting regimes in the dynamics of single-particle
correlations arising during the slow interaction quench of a
one-dimensional Bose gas. We proposed a generalized picture for the propagation of 
the correlation evolution front based on the counterpropagation
of entangled quasiparticle pairs moving at each point of time at their 
instanteneous velocity.
Therefore, the evolution front does not simply spread as a light-cone as found following
a sudden parameter change~\cite{cazalilla_quench,BarmettlerKollath2012}, but aquires a more complex functional form. 
We expect this picture to apply to other models and quench forms as the evolution front can be predicted 
from the sole knowledge of the quasiparticle velocity. For example, we expect that for
of a linear decrease of the interaction strength $U(t)=U_0(1-\frac{t}{t_0})$, starting from a Mott-insulator,
the propagation front will be of the form $4 J (2 n + 1)~t \left(1 - \frac{8 n (n + 1) J^2}{(2n + 1)^2 U_0^2 (1 - t/t_0)}\right)$
as the maximal velocity of quasiparticles is given by 
$v_{max}\approx 2J(2 n + 1) \left(1 - \frac{8n(n + 1)J^2}{(2n + 1)^2 U_0^2 (1 - t/t_0)^2}\right)$
where $n$ is the average filling~\cite{BarmettlerKollath2012}.
These results may serve as a basis for comparison with experimental studies of unconventional
time evolutions in many-body one-dimensional systems.

\acknowledgments We are grateful to T. Giamarchi for helpful discussions and to G. Roux for his insights on related works. 
We acknowledge support from ANR (FAMOUS), SNFS under MaNEP and Divison II, NSERC of Canada and CIFAR.


\clearpage


\newcommand{\eqthetacorrelator}{$8$}
\newcommand{\eqeomsolutiona}{$4$}
\newcommand{\eqeomsolutionb}{$5$}
\newcommand{\eqbesselramp}{$6$}
\renewcommand{\theequation}{S\arabic{equation}}

\setcounter{equation}{0}
\setcounter{page}{1}

\onecolumngrid

\vspace{0.5cm}
\begin{center}
\Large
Supplementary material
\end{center}

\section{Derivation of Eq.~(\eqthetacorrelator) of the main text}
Within the bosonization formalism the equal-time single-particle correlation function is   
\begin{eqnarray}
G(x,t)_{q\simeq 0} &=& A_0^2  \langle e^{i\theta(x,t)}  e^{-i\theta(0,t)}\rangle \nonumber \\
&=& A_0^2 e^{-\frac 1 2 \langle (\theta(x,t) -\theta(0,t))^2 \rangle}. 
\end{eqnarray}
The correlator appearing above is obtained from Eqs.~(\eqeomsolutiona) and (\eqeomsolutionb) of 
the main text and is given by
\begin{eqnarray}
\langle (\theta(x,t)-\theta(0,t)^2\rangle = \frac {1}{2 K_0 u_0^2}
\int_{-\infty}^{+\infty} \frac{dq}{|q|^3} e^{-\alpha |q|} \left|\frac{d}{dt}F(q,t)\right|^2 [1-\cos(qx)]
\end{eqnarray}
where the derivative of Eq.~(\eqbesselramp) of the main text is
\begin{eqnarray}
\frac{d}{dt}F(q,t) &=& \frac{\sqrt{3}}{2} \frac{u_0}{l_0}\pi \tau s^2
\left[J_\frac{2}{3}\left(s\right) J_{-\frac{2}{3}}(s \tau^\frac{3}{2} ) -
      J_{-\frac{2}{3}}\left(s\right)  J_\frac{2}{3}(s\tau^\frac{3}{2} ) \right.\nonumber + i
\left. \left(J_{-\frac{1}{3}}(s) J_{-\frac{2}{3}}(s \tau^\frac{3}{2} )+
       J_\frac{1}{3}\left(s\right)  J_\frac{2}{3} (s \tau^\frac{3}{2} ) \right)  \right]
 \end{eqnarray}
with $J_n(y)$ the Bessel function of the first kind. Recall that $l_0=u_0t_0$ is the typical lengthscale, 
$s(q,t_0)=\frac 2 3 l_0 |q| $ the dimensionless momentum, and $\tau(t,t_0)=1+\frac{t}{t_0}$ the dimensionless time.

\section{Asymptotic expansions}
In order to obtain analytical approximations for the single-particle function
$G(x,t)_{q\simeq 0}= A_0^2 e^{-\frac{1}{2} I(\xi,\tau,\tilde{\alpha})}$, we need to approximate the integral $I(\xi,\tau,\tilde{\alpha})$.
We identify three important regimes:  $s \ll \frac{2}{3} \tau^{-\frac{3}{2}}$, $ \frac{2}{3} \tau^{-\frac{3}{2}} \ll s \ll \frac{2}{3}$ 
and $s \gg \frac{2}{3}$ (note that $\tau\ge 1$ and that we separate the regimes in $s$ at $\frac 2 3$
instead of $1$ as this corresponds to a splitting at $\frac 1 l_0$ in momentum space). 
These three regimes determine the size of the arguments of the Bessel functions entering the 
integral $I(\xi,\tau,\tilde{\alpha})$. Therefore, we split this integral into three parts
$I(\xi,\tau,\tilde{\alpha}) = \mathfrak{G}\left( \xi,\tau \right) + \mathfrak{H}\left( \xi,\tau \right) + \mathfrak{Q}(\xi,\tau)$ with 
$\mathfrak{G}(\xi,\tau)=I(\xi,\tau,\tilde{\alpha}, 0, \frac{2}{3} \tau^{-\frac 3 2})$, 
$\mathfrak{H}(\xi,\tau)=I(\xi,\tau,\tilde{\alpha}, \frac 2 3 \tau^{-\frac 3 2}, \frac 2 3)$, and 
$\mathfrak{Q}(\xi,\tau)=I(\xi,\tau,\tilde{\alpha}, \frac 2 3,\infty)$ using 
\begin{eqnarray}
 I(\xi,\tau,\tilde{\alpha},c_1,c_2) =&& \frac{\pi^2 \tau^2}{3 K_0} \int_{c_1}^{c_2}
ds~s~e^{-\tilde{\alpha}s}~[1-\cos(s\xi)] \times \\
&& \left[\left(J_{\frac 2 3}(s) J_{- \frac 2 3}(s\tau^{\frac 3 2}) - J_{-\frac 2 3}(s) J_{\frac 2 3}(s \tau^{\frac 3 2})\right)^2 +
\left(J_{-\frac 1 3}(s) J_{-\frac 2 3}(s \tau^{\frac 3 2}) + J_{\frac 1 3}(s) J_{\frac 2 3}(s \tau^{\frac 3 2})\right)^2 \right]. \nonumber
\end{eqnarray}

The Bessel functions of the first kind can take simpler approximated forms
(using expressions taken from Ref.~\onlinecite{abramowitz_math_functions_supp}) in the two following regimes:
\begin{eqnarray}
\text{for}~s \ll \frac 2 3,~~J_{\nu}(s) &\sim& \left(\frac{s}{2}\right)^\nu \frac{1}{\Gamma(\nu+1)}, \\
\text{for}~s \gg \frac 2 3,~~J_{\nu}(s) &\sim& \sqrt{\frac 2 {\pi s}} \cos \left[ s - \frac{\pi} 4(2\nu+1)\right].
\end{eqnarray}

\subsection{Large momentum contribution}
For the function $\mathfrak{Q}(\xi,\tau)$ which covers the regime $s \gg \frac 2 3$ , we get
\begin{eqnarray}
\mathfrak{Q}(\xi,\tau) \simeq \frac{1}{K(\tau)} \int_{2/3}^{+\infty} \frac
{ds}{s} [1-\cos(s \xi)] e^{-s \tilde{\alpha}}=
\frac{1}{K(\tau)} \left\{
E_1\left(\frac 2 3 \tilde{\alpha}\right) 
- \frac 1 2 \left[E_1\left(\frac 2 3 (\tilde{\alpha}+ i\xi)\right) 
+ E_1\left(\frac 2 3 (\tilde{\alpha}- i\xi)\right)\right]\right\},
\end{eqnarray}
where $K(\tau) = \frac{K_0}{\sqrt{\tau}}$ and $E_1$ is the exponential integral
function~\cite{abramowitz_math_functions_supp}. For $\xi \ll 1$, we find
\begin{eqnarray}
\mathfrak{Q}(\xi,\tau)\simeq \frac{1}{2 K(\tau)}
\ln \left(1+\frac {\xi^2}{\tilde{\alpha}^2}\right),
\end{eqnarray}
while for $\xi \gg 1$,
\begin{eqnarray}
\mathfrak{Q}(\xi,\tau)\simeq \frac{1}{K(\tau)} \left[ \ln
\left(\frac { e^{-\gamma_E} }{\frac{2}{3} \tilde{\alpha}}\right) + \frac{1}
\xi \sin \left(\xi\right) + O(\xi^{-2}) \right]
\end{eqnarray}
where $\gamma_E$ is Euler's constant. 

\subsection{Small momentum contribution}
For the small momentum contribution, given by $\mathfrak{G}(\xi,\tau)$, we obtain the following expression
\begin{eqnarray}
\mathfrak{G}\left( \xi,\tau \right) &\simeq& 
\frac 1 {K_0} \int_0^{\frac{2\xi}{3} \tau^{-\frac 3 2}} \frac{du}{u} [1-\cos u] \nonumber \\ 
&\simeq& \frac 1 {K_0} \mathrm{Cin}\left(\frac{2\xi}{3} \frac{1}{\tau^{\frac 3 2}} \right).
\end{eqnarray}

\subsection{Intermediate momentum contribution}
The most interesting contribution is the intermediate momentum one which is due to $\mathfrak{H}(\xi,\tau)$. This expression simplifies to
\begin{eqnarray}
\mathfrak{H}\left( \xi,\tau \right) &\simeq&
\frac{2^{\frac 7 3} \pi \sqrt{\tau}}{3 K_0 (\Gamma(\frac 1 3))^2}~\Phi_1(\xi,\tau^{\frac 3 2}) + 
\frac{2^{\frac 5 3} \pi \sqrt{\tau}}{3 K_0 (\Gamma(\frac 2 3))^2}~\Phi_2(\xi,\tau^{\frac 3 2})
\end{eqnarray}
where
\begin{eqnarray}
\Phi_1(\xi, \tau^{\frac 3 2}) &=& \int_{\frac{2}{3} \tau^{-\frac 3 2}}^{\frac 2 3} \frac{ds}{s^{\frac 4 3}}~[1-\cos(s\xi)]~\sin^2(s\tau^{\frac 3 2}-\frac{\pi}{12}), \\
\Phi_2(\xi, \tau^{\frac 3 2}) &=& \int_{\frac{2}{3} \tau^{-\frac 3 2}}^{\frac 2 3} \frac{ds}{s^{\frac 2 3}}~[1-\cos(s\xi)]~\cos^2(s\tau^{\frac 3 2}+\frac{\pi}{12}). 
\end{eqnarray}

In the limit where $\tau \gg 1$ and $\xi \gg 1$, this expression can be further simplified and we find
\begin{eqnarray}
\mathfrak{H}\left( \xi,\tau \right) &\simeq&
\frac{2^{\frac 4 3} \pi^2 \tau^{\frac 1 2}}{K_0 \Gamma(\frac 1 3)^3}~\left(\xi^{\frac 1 3}- \frac{\Gamma(\frac 1 3)}{\pi} \left(\frac{3}{2}\right)^{\frac 1 3} \right)
+ \frac{2^{\frac 2 3}\pi \tau^{\frac 1 2}}{3~K_0 \Gamma(\frac 2 3)^2} 
\left(2^{\frac 1 3} 3^{\frac 2 3} - \frac{\pi^{\frac 1 2} \Gamma(\frac 1 6)}{2^{\frac 2 3} \Gamma(\frac 1 3)} \frac{1}{\xi^{\frac 1 3}}\right),
\end{eqnarray}
whereas in the limit where $\tau \gg 1$ and $\xi \ll 1$, we find
\begin{eqnarray}
\mathfrak{H}\left( \xi,\tau \right) &\simeq& 4 \pi \frac{\sqrt{\tau}}{K_0}~
\left(\frac{1}{5~3^{\frac 5 3} \Gamma(\frac 1 3)^2} + \frac{1}{7~3^{\frac 7 3} \Gamma(\frac 2 3)^2}\right)~\xi^2.
\end{eqnarray}

\subsection{Approximation for the full single-particle correlation function}
The full single-particle correlation function is given by  
$G(x,t)_{q\simeq 0}= A_0^2 e^{-\frac{1}{2} (\mathfrak{G}\left( \xi,\tau \right)+\mathfrak{H}\left( \xi,\tau \right)+\mathfrak{Q}(\xi,\tau))}$. 
We now use the expressions derived above to find approximations for the entire function.
 
\begin{itemize}
\item[(i)] Let us first discuss the limit of {\it large distances} such that $\xi \gg \tau^{\frac 3 2}$ (which 
necessarily implies that $\xi \gg 1$). In this limit, the large and intermediate momentum contributions 
become, to first order, independent of $\xi$ while the low momentum 
contribution $\mathfrak{G}(\xi,\tau) \simeq \frac 1 {K_0} \mathrm{Cin}\left(\frac{2\xi}{3} \tau^{-\frac 3 2} \right) \simeq \frac{1}{K_0} \left(\gamma_E + \ln(\xi) \right)$ dominates.
Thus the single-particle correlation behaves as
\begin{eqnarray}
G(\xi, \tau)_{q \simeq 0} &\simeq& \mathcal{A}(\tau) \left(\frac{1}{\xi}\right)^{\frac{1}{2 K_0}}
\end{eqnarray}
where $\mathcal{A}(\tau)$ is a function which depends on $\tau$.
We see that the Luttinger liquid behavior is recovered for long distances and that the exponent is the one expected in equilibrium before the ramp.

\item[(ii)] Let us now discuss the case of {\it small distances} $\xi \ll \tau^{-\frac 1 4}$.
In this limit, one obtains that the large momentum contribution, $\mathfrak{Q}(\xi,\tau)$, dominates since the intermediate 
and small momentum contribution vanish algebraically. Thus the correlation is well described by the expression
\begin{eqnarray}
G(\xi, \tau)_{q \simeq 0} \simeq \mathcal{B}(\tau) \left(1 + \frac{\xi^2}{\tilde{\alpha}^2}\right)^{-\frac{\sqrt{\tau}}{4 K_0}}
\end{eqnarray}
with $\mathcal{B}(\tau)$ a function depending on $\tau$.

\item[(iii)] For intermediate distances, finding an asymptotic expression for the single-particle correlation is a more subtle task. 
However, for $1 \ll \xi \ll \tau^{\frac 3 2}$, one obtains
\begin{eqnarray}
G(\xi, \tau)_{q \simeq 0} &\simeq& \mathcal{C}(\tau)
\exp\left[-\frac{2^{\frac 1 3} \pi^2 \tau^{\frac 1 2}}{K_0 \Gamma(\frac 1 3)^3}~\left(\xi^{\frac 1 3}- \frac{\Gamma(\frac 1 3)}{\pi} \left(\frac{3}{2}\right)^{\frac 1 3} \right)\right] \\
&&~~~~~~ \times \exp\left[-\frac{\pi \tau^{\frac 1 2}}{3~2^{\frac 1 3} K_0 \Gamma(\frac 2 3)^2} 
\left(2^{\frac 1 3} 3^{\frac 2 3} - \frac{\pi^{\frac 1 2} \Gamma(\frac 1 6)}{2^{\frac 2 3} \Gamma(\frac 1 3)} \frac{1}{\xi^{\frac 1 3}}\right)\right]. \nonumber
\end{eqnarray}
where $\mathcal{C}(\tau)$ a function that depends on $\tau$. This expression is due to the contribution $\mathfrak{H}(\xi, \tau)$ at intermediate momentum, and for sufficiently large $\xi$ the first exponential is dominant.
\end{itemize}

\section{Fourier transform of a stretched exponential}
The Fourier transform of a stretched exponential function of the form $f(x)=e^{-x^\beta}$ with $\beta<1$ is given by the expression
\begin{equation}
\mathcal{Q}(q) = \int_0^{\infty} dx e^{-x^\beta} e^{iq x}=\sum_{0}^{\infty}
\int_0^{\infty} \frac{(-x^\beta)^n}{n!}e^{iq x} =\sum_{0}^{\infty}
\frac{(-1)^n}{n!} \int_0^{\infty} du \frac{u^{\beta n}
  e^{iu}}{q^{\beta n+1}}=\sum_{0}^{\infty}
\frac{(-1)^n}{\Gamma(n+1)}\frac{\Gamma(\beta n+1)}{q^{\beta
    n+1}}e^{i\frac{\pi}{2}(\beta n+1)}
\end{equation}
which is a convergent sum. When $\beta=\frac 1 3$, the sum can be further evaluated by using
properties of the $\Gamma$ functions, $\frac{\Gamma(\frac{n}{3}+1)}{\Gamma(n+1)}=\frac{1}{3}\frac{\Gamma(\frac{n}{3})}{\Gamma(n)}$,
and definitions of the Hypergeometric functions~\cite{abramowitz_math_functions_supp}, this exercise yields the result
\begin{eqnarray}
\mathcal{Q}(q)=\frac{e^{-i \frac \pi 2 \mathrm{sign}(q)}}{|q|} \left[ 1 -
      \frac{\pi e^{-i \frac \pi 6 \mathrm{sign}(q) }}{(3|q|)^{\frac 1 3}}
      \mathrm{Hi}\left(\frac{-e^{-i \frac \pi 6
      \mathrm{sign}(q)}}{(3|q|)^{\frac 1 3}}\right)\right]
\end{eqnarray}
where $\mathrm{Hi}$ is the Scorer function~\cite{abramowitz_math_functions_supp,gil2001,nikishov2005}.


\begin{thebibliography}{46}
\bibitem{footnote}
Equivalently, the solution can be expressed using Airy
functions, but the Bessel function representation leads to a more straightforward derivation 
of asymptotics.
\end{thebibliography}

\end{document}